\documentclass[a4paper]{PoS}

\title{Loop calculations for models of graphene}

\ShortTitle{Loops for graphene}

\author{\speaker{J.A. Gracey}\\
        Theoretical Physics Division, Department of Mathematical Sciences,
        University of Liverpool, P.O. Box 147, Liverpool, L69 3BX,
        United Kingdom \\
        E-mail: \email{gracey@liverpool.ac.uk}}

\abstract{The various Gross-Neveu classes of quantum field theories are of 
interest in the study of a particular phase transition in graphene. We review 
recent activity in the determination of several critical exponents in the 
chiral Heisenberg Gross-Neveu universality class using the critical point large
$N$ technique which includes the use of the large $N$ conformal bootstrap 
method.}

\FullConference{Loops and Legs in Quantum Field Theory (LL2018)\\
		29 April 2018 - 04 May 2018\\
		St. Goar, Germany}

\newcommand{\xslash}{x \! \! \! /}

\newcommand{\partialslash}{\partial \! \! \! \! \! /}
\newcommand{\half}{\mbox{\small{$\frac{1}{2}$}}}

\begin{document}

\section{Introduction}

The Gross-Neveu model, \cite{1}, has been of interest in recent years due to 
its connection with a phase transition in graphene. This is a one atom thick 
layer of carbon atoms arranged in a honeycomb lattice. In particular when the 
carbon sheet is deformed or stretched its electrical conductivity properties 
can change from a conductor to a Mott-insulating phase, \cite{2,3}. The 
critical dynamics of this phase transition is believed to be described by the 
chiral Heisenberg Gross-Neveu (cHGN) model, \cite{4,5,6,7,8}. As the 
fundamental interaction of this and the other Gross-Neveu field theories is a 
quartic fermion self-interaction there is also a connection with the Standard 
Model of particle physics. At the Wilson-Fisher fixed point the ultraviolet 
completion of the Gross-Neveu model in four dimensions is the
Gross-Neveu-Yukawa theory (GNY), \cite{9}. In this formulation the auxiliary 
scalar field of the two dimensional theory is transformed into a propagating
fundamental field with a quartic scalar interaction in addition to the Yukawa 
one. Structurally the basic Lagrangian is similar to the Standard Model. 
Moreover prior to the discovery of the Higgs particle such Gross-Neveu-Yukawa 
models were used to explore the idea that the Higgs particle was a composite of 
fermions. For example, see \cite{10}. 

The current interest in the application to graphene centres on the 
determination of the critical exponents governing the phase transition. In
general these can be computed for the physically relevant case of three
dimensions by using techniques such as the $\epsilon$ expansion, Functional
Renormalization Group (FRG), Monte Carlo methods or the large $N$ expansion. 
While the Ising Gross-Neveu model has been extensively examined in perturbation
theory as well as the other techniques to very high precision with in general 
very solid agreement across all methods, the status of the more interesting 
chiral Heisenberg Gross-Neveu universality class is not yet at the same level. 
For instance, the Functional Renormalization Group study of \cite{8} is 
primarily the first reliable field theory examination of the exponents $\eta$, 
$\eta_\phi$ and $\nu$. Indeed Figures $1$, $2$ and $3$ of \cite{8} reflect 
their status. However since that article there has been significant work in the
four dimensional ultraviolet completion at four loops, \cite{11,12,13}. Also 
the same critical exponents have been determined to three orders in the large 
$N$ expansion, \cite{14}. We will therefore review recent progress of the 
latter in this area. 

\section{Background}

To appreciate the connection of the various theories it is best to recall the
basic or Ising Gross-Neveu model which has the Lagrangian, \cite{1},
\begin{equation}
L^{\mbox{\footnotesize{GN}}} ~=~ i \bar{\psi}^i \partialslash \psi^i ~-~
m \bar{\psi}^i \psi^i ~+~ g \tilde{\sigma} \bar{\psi}^i \psi^i  ~-~
\frac{1}{2} \tilde{\sigma}^2
\label{laggn}
\end{equation}
which is renormalizable in two dimensions and $1$~$\leq$~$i$~$\leq$~$N$. Here 
the scalar field $\sigma$ is auxiliary and if eliminated produces a theory with
a quartic fermion self-interaction. The connection with the Gross-Neveu-Yukawa
model is apparent from the Lagrangian, \cite{9}, 
\begin{equation}
L^{\mbox{\footnotesize{GNY}}} ~=~ i \bar{\psi}^i \partialslash \psi^i ~-~
m \bar{\psi}^i \psi^i ~+~
\frac{1}{2} \partial_\mu \tilde{\sigma} \partial^\mu \tilde{\sigma} ~+~
\frac{1}{2} g_1 \tilde{\sigma} \bar{\psi}^i \psi^i ~+~
\frac{1}{24} g_2^2 \tilde{\sigma}^4
\label{laggny}
\end{equation}
which is renormalizable in four dimensions. In this case the scalar field 
propagates as a fundamental field and the quartic scalar interaction is 
necessary to ensure renormalizability. The common scalar-fermion interaction
shared in both Lagrangians is what underlies the universality and leads to 
critical point equivalence at the $d$-dimensional Wilson-Fisher fixed point. 
While (\ref{laggny}) has now been renormalized in four dimensions to four loops
in \cite{11,12,13} the $\epsilon$ expansion extrapolation of critical exponents
to three dimensions has been carried out in \cite{13} using matched Pad\'{e} 
approximants. In this a Pad\'{e} estimate is constructed using data from 
{\em two} and {\em four} dimensional perturbation theory. A recent 
comprehensive analysis collating data from all approaches has been discussed in
\cite{15}. To facilitate such a construction the four loop $\beta$-function of 
(\ref{laggn}) was provided in \cite{16} as  
\begin{eqnarray}
\beta(g) &=& (d-2)g \,-\, ( N - 1 ) \frac{g^2}{\pi} ~+~
( N - 1 ) \frac{g^3}{2\pi^2} ~+~ ( N - 1 ) ( 2N - 7 ) \frac{g^4}{16\pi^3}
\nonumber \\
&& +~ ( N - 1 ) \left[ - 2 N^2 - 19 N + 24 - 6 \zeta_3 ( 11 N - 17 )
\right] \frac{g^5}{48\pi^4} ~+~ O(g^6) 
\end{eqnarray}
where $\zeta_z$ is the Riemann zeta function. With this four loop
$\beta$-function, estimates of the three exponents were provided in \cite{16} 
for three dimensions which were commensurate with other techniques.

While this Ising Gross-Neveu universality class has been studied in depth a
more apt one for the graphene phase transition of interest is the chiral
Heisenberg Gross-Neveu universality class. Its two dimensional Lagrangian is
\begin{equation}
L^{\mbox{\footnotesize{cHGN}}} ~=~ i \bar{\psi}^i \partialslash \psi^i ~-~
m \bar{\psi}^i \psi^i ~+~ g \tilde{\pi}^a \bar{\psi}^i \sigma^a \psi^i  ~-~
\frac{1}{2} \tilde{\pi}^a \tilde{\pi}^a
\label{lagchgn}
\end{equation}
where $\sigma^a$ are the three Pauli spin matrices. The four dimensional
ultraviolet completion is the chiral Heisenberg Gross-Neveu-Yukawa (cHGNY)
model which has the Lagrangian 
\begin{equation}
L^{\mbox{\footnotesize{cHGNY}}} ~=~ i \bar{\psi}^i \partialslash \psi^i ~-~
m \bar{\psi}^i \psi^i ~+~
\frac{1}{2} \partial_\mu \tilde{\pi}^a \partial^\mu \tilde{\pi}^a ~+~
\frac{1}{2} g_1 \tilde{\pi}^a \bar{\psi}^i \sigma^a \psi^i ~+~
\frac{1}{24} g_2^2 \left( \tilde{\pi}^a \tilde{\pi}^a \right)^2
\label{lagchgny}
\end{equation}
parallel to (\ref{laggny}). At present the major activity in evaluating the
relevant critical exponents in (\ref{lagchgny}) has been with Functional
Renormalization Group and perturbative methods, \cite{4,5,6,7,8}. Given the 
complementarity of the large $N$ technique with both approaches we note that
the $1/N$ formalism is developed from the underlying universal Lagrangian 
\begin{equation}
L ~=~ i \bar{\psi}^i \partialslash \psi^i ~+~
\pi^a \bar{\psi}^i \sigma^a \psi^i  ~+~ f(\pi^a) ~.
\label{lagchgnun}
\end{equation}
Here $f(\pi^a)$ represents all additional operators, a small subset of which 
are relevant to defining a renormalizable Lagrangian in each even critical 
dimension.

\section{Large $N$ expansion}

We briefly summarize the approach of the critical point large $N$ method
developed for scalar models in \cite{17,18,19}. The key is that in the approach
to criticality the propagators satisfy a power law structure where the exponent
is comprised of a canonical and anomalous part. The former is deduced from a 
dimensional analysis of the $d$-dimensional Lagrangian and the latter reflects 
the radiative corrections. Both are functions of $d$~$=$~$2\mu$ with the 
anomalous part depending on $N$. At criticality $1/N$ plays the role of the 
dimensionless coupling constant. Specifically the coordinate space propagators 
take the scaling forms, \cite{20,21,22,23,24,25,26},
\begin{equation}
\psi(x) ~\sim~ \frac{A\xslash}{(x^2)^\alpha} \left[ 1 + A^\prime(x^2)^\lambda
\right] ~~~,~~~
\pi(x) ~\sim~ \frac{C}{(x^2)^\gamma} \left[ 1 + C^\prime(x^2)^\lambda
\right]
\end{equation}
where corrections to scaling are included through the exponent $\lambda$. This
could be the exponent of the critical $\beta$-function slope for either the
chiral Heisenberg Gross-Neveu or chiral Heisenberg Gross-Neveu-Yukawa models. 
For instance the critical slope of the $\beta$-function of (\ref{laggny}) was 
recently determined in \cite{27} to $O(1/N)$ and to the next order in 
\cite{28}. The exponents of the fields are given by
\begin{equation}
\alpha ~=~ \mu ~+~ \half \eta ~~~,~~~ \gamma ~=~ 1 ~-~ \eta ~-~ \chi_\pi
\end{equation}
where $\chi_\pi$ is the exponent of the universal vertex. For instance, the 
$1/N$ expansion of $\eta$ is defined by
\begin{equation}
\eta(\mu) ~=~ \sum_{n=1}^\infty \frac{\eta_n(\mu)}{N^n} ~.
\end{equation}
For (\ref{laggny}) the exponents $\eta$, $\chi_\pi$ and 
$\lambda$~$=$~$1/(2\nu)$ can be found to $O(1/N^2)$ in $d$-dimensions by 
algebraically solving the skeleton Schwinger-Dyson equations when the scaling 
forms of the propagators are used, \cite{17,18}. This involves using the 
so-called master integrals given in \cite{20,25} but decorating them with the 
group theory factors derived from the Pauli matrices. It is therefore 
straightforward to deduce the exponents  
\begin{eqnarray}
\eta_1 &=& -~ \frac{3 \Gamma(2\mu-1)}{\mu\Gamma(1-\mu)\Gamma(\mu-1)
\Gamma^2(\mu)} ~~~,~~~ 
\chi_{\pi\,1} ~=~ -~ \frac{\mu}{3(\mu-1)} \eta_1 \nonumber \\
\lambda_1 &=& -~ (2\mu-1) \eta_1 ~~~,~~~
\eta_2 ~=~ \left[ \frac{(2\mu-3)}{3(\mu-1)} \Psi(\mu) ~+~
\frac{(4\mu^2-6\mu+1)}{2\mu(\mu-1)^2} \right] \eta_1^2 
\end{eqnarray}
where
\begin{equation}
\Psi(\mu) ~=~ \psi(2\mu-1) ~-~ \psi(1) ~+~ \psi(2-\mu) ~-~ \psi(\mu) 
\end{equation}
and $\psi(z)$ is the polygamma function. An expression for $\nu$ at next order 
requires the evaluation of several four loop graphs with the correction to 
scaling included in a $\pi^a$ line. The result is more involved since  
\begin{eqnarray}
\lambda_2 &=& \left[ \frac{\mu^2[6\mu^2-3\mu-8]}{6(\mu-1)(\mu-2)}
\Theta(\mu) ~-~ \frac{2\mu^2(2\mu-3)}{3(\mu-1)(\mu-2)} \left[
\Phi(\mu) + \Psi^2(\mu) \right] \right. \nonumber \\
&& \left. ~+~
\frac{2\mu[8\mu^2-16\mu+7]}{3(\mu-1)(\mu-2)^2\eta_1} ~+~
\frac{[72 \mu^8 - 604 \mu^7 + 1960 \mu^6 - 3060 \mu^5]}
{18 \mu (\mu - 1)^3 (\mu - 2)^2}
\right. \nonumber \\
&& \left. ~+~
\frac{[2151 \mu^4 - 146 \mu^3 - 621 \mu^2 + 288 \mu - 36]}
{18 \mu (\mu - 1)^3 (\mu - 2)^2}
\right. \nonumber \\
&& \left. ~-~ \frac{(2\mu-3)[18\mu^5-95\mu^4+161\mu^3-86\mu^2-20\mu+12]}
{9(\mu-1)^2(\mu-2)^2} \Psi(\mu) \right] \eta_1^2 
\end{eqnarray}
with 
\begin{equation}
\Theta(\mu) ~=~ \psi^\prime(\mu) ~-~ \psi^\prime(1) ~~~,~~~ 
\Phi(\mu) ~=~ \psi^\prime(2\mu-1) ~-~ \psi^\prime(2-\mu) ~-~
\psi^\prime(\mu) ~+~ \psi^\prime(1) ~. 
\end{equation}
Given that the main computational tool to evaluate the large $N$ graphs is the
$d$-dimensional conformal integration or uniqueness method of \cite{16,17}, but
extended to the Gross-Neveu universality class in \cite{20}, it is possible to 
extend $\eta$ to the next order. This requires use of the large $N$ conformal 
bootstrap programme introduced in \cite{18} and then developed for
(\ref{laggn}) in \cite{23,26,29}. Application of the method leads to 
\begin{eqnarray}
\eta_3 &=& \left[ \frac{(2\mu-3)}{18(\mu-1)^2}
\left[ \Phi(\mu) + 3 \Psi^2(\mu) \right] ~-~
\frac{[\mu^3+18\mu^2-21\mu+9]}{36(\mu-1)^2} \Theta(\mu) \right. \nonumber \\
&& \left. ~-~ \frac{\mu^2}{3(\mu-1)} \Theta(\mu) \Psi(\mu) ~-~
\frac{\mu^2}{6(\mu-1)^3} \Xi(\mu) ~-~
\frac{\mu^2}{6(\mu-1)} \Xi(\mu) \Theta(\mu) \right. \nonumber \\
&& \left. ~-~
\frac{[14 \mu^7 - 15 \mu^6 - 26 \mu^5 - 77 \mu^4 + 324 \mu^3 - 297 \mu^2
+ 90 \mu - 9]}{18\mu^2(\mu-1)^4} \right. \nonumber \\
&& \left. ~-~ \frac{[14\mu^5-37\mu^4-50\mu^3+228\mu^2-183\mu+27]}
{18\mu(\mu-1)^3} \Psi(\mu)
\right] \eta_1^3 
\end{eqnarray}
where $\Xi(\mu)$ is related to a two loop self-energy Feynman graph, \cite{18}.

{\begin{table}[ht]
\begin{center}
\begin{tabular}{|c||l|l|l|l|}
\hline
$N$~$=$~$4$  & $1/\nu$ & $\eta_\phi$ & $\eta$ & $\nu$ \\
\hline
$\epsilon$ expansion $[2,2]$ \cite{13} &
$0.6426$ & $0.9985$ & $0.1833$ & $-$ \\
$\epsilon$ expansion $[3,1]$ \cite{13} &
$0.6447$ & $0.9563$ & $0.1560$ & $1.2352$ \\
FRG \cite{30} & $0.795$ & $1.032$ & $0.071$ & $1.26$ \\
MC \cite{31} & $(0.98)$ & & $0.20(2)$ & $1.02(1)$ \\
MC \cite{32} & $(1.19)$ & $0.70(15)$ & & $0.84(4)$ \\
Large $N$ & $0.8458$ & $1.1849$ & $0.1051$ & $1.1823$ \\
\hline
\end{tabular}
\end{center}
\end{table}}

\section{Results}

The main results of \cite{14} were the determination of the three key critical
exponents for the chiral Heisenberg Gross-Neveu universality class. A very
strong check on their correctness rests in the fact that if one expands each
expression in powers of $\epsilon$ where $d$~$=$~$4$~$-$~$2\epsilon$ and 
compares with the same expansion of the critical exponents derived from the 
four dimensional renormalization group functions of \cite{11,12,13}, then there
is {\em exact} agreement. One advantage of the large $N$ approach is that three
dimensional information can be immediately determined from the $d$-dimensional 
expressions. In particular we have  
\begin{eqnarray}
\eta &=& \frac{4}{\pi^2N} ~+~ \frac{64}{3\pi^4N^2} ~+~ 
\frac{8 [ 378 \zeta_3 - 36 \pi^2 \ln(2) - 45 \pi^2 - 332 ]}{9\pi^6N^3} ~+~ 
O \left( \frac{1}{N^4} \right) \nonumber \\
\eta_\phi &=& 1 ~+~ \frac{16[3\pi^2+16]}{3\pi^4N^2} ~+~ 
O \left( \frac{1}{N^3} \right) \nonumber \\
\frac{1}{\nu} &=& 1 ~-~ \frac{16}{\pi^2N} ~+~
\frac{16 [ 144 \pi^2 + 1664 ]}{3\pi^4N^2} ~+~ O \left( \frac{1}{N^3} \right)
\end{eqnarray}
where there is no $O(1/N)$ correction for $\eta_\phi$ which is related to
$\eta$~$+$~$\chi_\pi$. As the specific value of $N$ for graphene applications
is $N$~$=$~$4$ in our conventions we summarize the current status of exponent
estimates in Table $1$. Overall the large $N$ and Functional Renormalization
Group estimates are broadly similar with $\epsilon$ summation being less
consistent. There have only been a few Monte Carlo (MC) evaluations. In 
comparison with the status of the Ising Gross-Neveu class for the comparable
value of $N$ for graphene there is less close agreement between different
methods for the chiral Heisenberg Gross-Neveu universality class, \cite{13,14}.

\section{Discussion}

We have summarized the application of the $d$-dimensional critical point large 
$N$ method to continuum quantum field theories which are believed to underlie a 
specific phase transition in graphene. For the Ising Gross-Neveu universality
class a wide body of methods has produced critical exponent estimates which are
commensurate. For the more interesting chiral Heisenberg Gross-Neveu class the
picture is currently not as clear. This is in part due to the fact that there 
has only been activity on this class in more recent years. So the level of 
precision is not comparable with the Ising class. However the large $N$ 
exponents are now available to the same order as well as the four dimensional 
four loop exponents from \cite{8,13} while the FRG results were in effect the 
first $d$-dimensional exploration of this class. It is therefore clear that 
now effort should be devoted to bringing estimates from other methods to the 
same level as the Ising class. Aside from numerical Monte Carlo techniques the
determination of the four loop renormalization group functions for the two
dimensional chiral Heisenberg Gross-Neveu theory (\ref{lagchgn}) would produce 
matched Pad\'{e} approximants of the $d$-dimensional behaviour of the 
exponents.

\section*{Acknowledgements} This work was carried out with the support in part 
of a DFG Mercator Fellowship. Computations were carried out in part using the 
symbolic manipulation language {\sc Form}, \cite{33,34}.

\end{document}